\documentclass[reprint,amsmath,amssymb,amsart,aps,superscriptaddress,twocolumn,float]{revtex4-2}
\usepackage{graphicx,graphics}
\usepackage{dcolumn}
\usepackage{bm}
\usepackage{amssymb}
\usepackage{amsfonts}
\usepackage{float}
\usepackage{xcolor}
\usepackage{hyperref}

\hypersetup{
	pdfnewwindow=true,      
	colorlinks=true,       
	linkcolor=blue,          
	citecolor=blue,        
	filecolor=blue,      
	urlcolor=blue,          
}

\begin{document}

\title{Ab initio design of enhanced and sign-reversible spin Hall and spin Nernst conductivity in skutterudite materials}

\author{Saikat Debnath}
\email{sktdebnath@gmail.com}
\affiliation{M. V. College, Buxar, Bihar, India\looseness=-1}

\author{Babu Baijnath Prasad} 
\email{babubaijnathprasad@gmail.com}
\affiliation{Institute for Solid State Physics, The University of Tokyo, Kashiwa, Chiba 277-8581, Japan\looseness=-1}

\author{Shishir Kumar Pandey}
\email{shishir.kr.pandey@gmail.com}
\affiliation{Department of General Sciences, Birla Institute of Technology and Science, Pilani-Dubai Campus, Dubai International Academic City, Dubai 345055, UAE\looseness=-1}
\affiliation{Department of Physics, Birla Institute of Technology and Science, Pilani, Hyderabad Campus, Jawahar Nagar, Kapra Mandal, Medchal District, Telangana 500078, India}

\date{\today}

\begin{abstract}

The search for novel materials capable of robust spin-current generation remains an active area of research. Here, we investigate the intrinsic spin Hall and spin Nernst effects in Co- and Rh-based skutterudites and explore their enhancement through heavy-metal substitution and site-selective doping. Using first-principles calculations and Wannier-based tight-binding models, we analyze the spin Berry curvature and spin-transport responses of pristine MX$_3$, where M = Co, Rh and X = As, Sb, as well as chemically modified systems. Despite sizable momentum-resolved spin Berry curvature, the pristine compounds exhibit relatively small net spin Hall responses due to substantial cancellation across the Brillouin zone. Isoelectronic Ir and Bi substitution at 8$c$ Wycoff M site produces little enhancement near the E$_F$, whereas non-isoelectronic Pt substitution yields a relatively better response. Most interestingly, the incorporation of Pt at the interstitial $2a$ Wycoff site of RhAs$_3$ nearly doubles the spin Hall conductivity to $\sim400~(\hbar/e)(\Omega,\mathrm{cm})^{-1}$ near the E$_F$ when compared to Pt substitution at 8$c$ position. At 300 K, the same interstitially filled system exhibits a spin Nernst conductivity of $\sim1.46~(\hbar/e),\mathrm{A,m^{-1}K^{-1}}$ at the E$_F$, nearly an order of magnitude larger than that of $M$-site Pt substitution. The enhanced responses arise from Pt $5d$ states, strong spin-orbit coupling, and avoided crossings that generate pronounced spin-Berry-curvature hot spots. These results demonstrate that site-selective chemical engineering can effectively overcome momentum-space cancellation resulting in enhanced spin Hall and spin Nernst responses in skutterudites. Our results provide a possible route towards robust spin-current generation in this class of materials.

\end{abstract}

\maketitle

\section{INTRODUCTION}

Efficient interconversion among charge, heat, and spin currents is an important objective in spintronics and spin caloritronics. In a nonmagnetic solid with spin-orbit coupling (SOC), an electric field can produce a spin current transverse to the direction of charge transport through the spin Hall effect (SHE), originally proposed in 1971 by D'yakonov and Perel~\cite{Dyakonov1971}. Because this effect does not rely on magnetic order or an externally applied magnetic field, it is particularly attractive for electrical generation of spin currents~\cite{Hirsch1999,Murakami2003,Sinova2004,Kato2004,Guo2005,Valenzuela2006}. Heavy transition metals provide prominent realizations of this phenomenon: Pt, for example, exhibits a large intrinsic spin Hall response arising from its strong SOC and multiband electronic structure~\cite{Kimura2007,Guo2008,Tanaka2008}. A temperature gradient can generate an analogous transverse spin current through the spin Nernst effect (SNE)~\cite{Cheng2008}. The SNE therefore extends charge-to-spin conversion to a thermal driving force and offers a means of utilizing heat for spin-current generation~\cite{Bauer2012,Wimmer2013}. Experimental observations in the heavy metals Pt and W have demonstrated that appreciable spin Nernst responses can occur in materials with strong SOC~\cite{Meyer2017,Sheng2017}. The SHE and SNE consequently provide complementary electrical and thermal routes for producing spin currents in nonmagnetic systems.

These considerations have stimulated searches for intrinsic SHE and SNE in materials with nontrivial band structures~\cite{Okuma2016,Sun2016,Sun2017,MacNeill2017,Shi2019,Yen2020,Zhao2020,Prasad2020,Hsieh2022,Golovanova2023,Ascencio2023}. In particular, SOC-gapped nodal features and chiral multifold fermions have been shown to generate sizable and tunable spin Hall and spin Nernst responses. These results demonstrate that crystalline symmetry, orbital hybridization, and the position of the chemical potential can jointly control intrinsic spin-current generation. This motivates the search for material families in which these factors can be systematically tuned by chemical composition. Binary transition-metal pnictide skutterudites provide a suitable platform for this purpose. They have the general formula \(MX_3\), where \(M\) is a transition-metal element and \(X\) is a pnictogen; in the present work, we focus on the Co- and Rh-based arsenides and antimonides CoAs\(_3\), CoSb\(_3\), RhAs\(_3\), and RhSb\(_3\). These compounds crystallize in the cubic skutterudite structure with space group \(Im\bar{3}\) (No.~204), in which the transition-metal atoms occupy the $8c$ \((1/4,1/4,1/4)\) Wyckoff sites and the pnictogen atoms occupy the $24g$ \((0,y,z)\) sites~\cite{Luo2015}. The skutterudite framework can be described in terms of distorted \(MX_6\) octahedra and four-membered pnictogen rings, with two empty cages centered at the $2a$ Wyckoff position \((0,0,0)\) per conventional unit cell~\cite{Anno2000,Luo2015}. This framework, together with the possibility of cage filling and framework-site substitution, has made skutterudites a versatile platform for thermoelectric materials design~\cite{Fleurial1994,Luo2015}. In particular, filler atoms and alloy disorder can reduce lattice thermal conductivity, while carrier doping and site substitution provide routes to tune the electronic transport properties~\cite{Watcharapasorn2000,Stoica2012,Luo2015,Sun2015}.

Photoemission measurements and electronic-structure calculations on Co- and Rh-based arsenide and antimonide skutterudites indicate appreciable transition-metal-\(d\)/pnictogen-\(p\) mixing in CoAs\(_3\), CoSb\(_3\), and RhSb\(_3\), together with a low density of states near the E$_F$ associated with a dispersive band in the pseudogap region~\cite{Anno2000}. Such hybridized near-Fermi states are relevant for the SHE and SNE because SOC-induced band splittings and interband matrix elements can strongly influence the spin Berry curvature (SBC). In addition, skutterudite-structure pnictides have been discussed in the context of unusual low-energy linear bands and topological transitions involving zero-momentum Dirac- or Weyl-type semimetallic states~\cite{Smith2011,Pardo2012}. RhAs\(_3\) is particularly relevant in this regard, as it has been identified as a candidate quadratic-contact-point semimetal, with a triply degenerate contact point in the absence of SOC that becomes fourfold degenerate when SOC is included~\cite{Zhu2018}. Although skutterudites have been extensively studied for their structural, electronic, and thermoelectric properties, their intrinsic spin Hall and spin Nernst responses have not yet been investigated. Since the spin Hall conductivity (SHC) and spin Nernst conductivity (SNC) are governed by spin-current matrix elements and the momentum-space distribution of spin Berry curvature, they need not follow trends in conventional electrical conductivity or Seebeck coefficient. The compounds CoAs\(_3\), CoSb\(_3\), RhAs\(_3\), and RhSb\(_3\) therefore provide a useful platform to examine how transition-metal choice, pnictogen chemistry, and chemical-potential positioning influence the relativistic band structure and intrinsic spin-transport coefficients.

Therefore, in this work, we investigate the intrinsic SHE and SNE of the above mentioned materials using relativistic density-functional theory and Wannier-function-based tight-binding (WFTB) models. We calculate the SHC over a wide chemical-potential range and analyze the momentum- and band-resolved spin Berry curvature to identify the electronic origin of the dominant contributions. The SNC is then evaluated from the energy-dependent SHC at finite temperature, enabling a direct comparison between electrically and thermally driven spin-current responses. We find that the pristine compounds exhibit relatively small net SHC despite sizable momentum-resolved SBC, owing to substantial cancellation of positive and negative contributions across the Brillouin zone. To overcome this inherent cancellation, we further examine site-selective incorporation of heavy $5d$ elements (Pt and Ta) as a strategy to tune these responses. By contrasting substitutional doping at the 8$c$ Wycoff \(M\)-site with interstitial filling at the empty $2a$ Wycoff site, we demonstrate how dopant species, valence electron count, and atomic site selection modify orbital hybridization, and unmask SBC hot spots. This may provide a robust strategy to achieve highly enhanced SHC and SNC in skutterudite-based materials.

The remainder of the paper is organized as follows. Section II details the theoretical framework and computational methods. Section III-A analyzes the electronic structure of the pristine compounds. Section III-B evaluates the spin Hall response of these pristine materials. Section III-C explores the spin Hall response under isoelectronic heavy element substitution. Section III-D examines the role of site-selective doping in enhancing the spin Hall response and analyzes the corresponding spin Nernst conductivity and its temperature dependence. Finally, Section IV provides a summary of our results.

\section{THEORY \& COMPUTATIONAL DETAILS}

{\it Ab initio} electronic structure calculations for skutterudites $\text{MX}_3$ (where M = Rh or Co; X = As or Sb) were performed using Density Functional Theory (DFT) as implemented in the Quantum ESPRESSO suite~\cite{Giannozzi2009,Giannozzi2017,Giannozzi2020}. The exchange-correlation functional was treated within the generalized gradient approximation (GGA) parameterized by Perdew, Burke, and Ernzerhof (PBE)~\cite{Perdew1996}. 
To accurately capture relativistic effects in heavy transition-metal and pnictogen elements, we employed fully relativistic, spin-orbit-coupled projector-augmented-wave (PAW) pseudopotentials, with SOC explicitly included within the non-collinear formulation of Quantum ESPRESSO.

The electronic wavefunctions were expanded in a plane-wave basis set with a minimum kinetic energy cutoff of $50.0\text{ Ry}$, while the charge density and potential evaluations were supported by an augmented density cutoff of minimum $400.0\text{ Ry}$. 
The self-consistent field (SCF) equations were solved iteratively with a tight convergence threshold of $1.0 \times 10^{-10}\text{ Ry}$. 
Brillouin zone integrations over the self-consistent cycle were sampled using a Monkhorst-Pack automatic $\mathbf{k}$-point mesh of $8 \times 8 \times 8$. 
To facilitate electronic convergence near the E$_F$, a Marzari-Vanderbilt cold smearing scheme was introduced with a broadening width of $0.01\text{ Ry}$. The experimental skutterudite structure contains 32 atoms in the body-centered cubic conventional unit cell. By exploiting the crystal symmetry, we reduced this to the corresponding $16$ atom primitive cell. This structure was used for the
pristine-structure calculations (see Table~\ref{tab:lattice_constants}). After site substitutions, the lattice parameters as well as the internal coordinates were fully relaxed to minimize the internal stress of the system. 
 For interstitial filling at the $2a$ Wyckoff position, the unit cell contains seventeen atoms.

To precisely evaluate the spin Hall response over an ultra-fine mesh, full relativistic {\it ab initio} band structures were mapped onto WFTB as implemented in Wannier90 code~\cite{Marzari1997,Souza2001,Mostofi2014}.
An active subspace comprising atomic projections from M $d$-orbitals and X $p$-orbitals was used. 
The tight-binding models were used to evaluate the intrinsic SHC using the postw90 module of Wannier90~\cite{Pizzi2020}.

Specifically, the intrinsic spin Hall conductivity is computed using the Kubo formula expressed in terms of the spin Berry curvature~\cite{Guo2008,Xiao2010}.
For a spin current flowing along the $i$ direction with spin polarization $s$ in response to an electric field applied along the $j$ direction, the SHC tensor component $\sigma_{ij}^{s}$ is written as~\cite{Guo2008},
\begin{equation}
\label{eq:shc}
\begin{aligned}
\sigma_{ij}^{s}
=
e\sum_n\int_{\rm BZ}\frac{d{\bf k}}{(2\pi)^3}
f_{{\bf k}n}\Omega_{ij}^{n,s}({\bf k}),
\end{aligned}
\end{equation}
where $f_{{\bf k}n}$ is the Fermi--Dirac distribution function for band $n$ at wave vector ${\bf k}$. The band-resolved spin Berry curvature is given by
\begin{equation}
\label{eq:sbc}
\begin{aligned}
\Omega_{ij}^{n,s}({\bf k})
=
\sum_{n'\neq n}
\frac{
2{\rm Im}\left[
\langle{\bf k}n|\{\tau_s,v_i\}/4|{\bf k}n'\rangle
\langle{\bf k}n'|v_j|{\bf k}n\rangle
\right]
}
{(\epsilon_{{\bf k}n}-\epsilon_{{\bf k}n'})^2+\eta^2}.
\end{aligned}
\end{equation}
Here $v_i$ and $v_j$ are velocity operators, $\tau_s$ is the Pauli matrix corresponding to the spin-polarization direction $s$, and $\eta$ is a small broadening parameter introduced in the energy denominator. 
After obtaining the spin Hall conductivity as a function of energy, the intrinsic spin Nernst conductivity is calculated as~\cite{Xiao2010},
\begin{equation}
\label{eq:snc}
\begin{aligned}
\alpha_{ij}^{s}
=
\frac{1}{e}\int_{-\infty}^{\infty}d\varepsilon
\frac{\partial f}{\partial\varepsilon}
\sigma_{ij}^{s}(\varepsilon)
\frac{\varepsilon-\mu}{T},
\end{aligned}
\end{equation}
where $\mu$ is the chemical potential and $T$ is the temperature. 
The energy-dependent SNC is evaluated at $T=300$ K, while its temperature dependence is calculated at the E$_F$.

Furthermore, to systematically observe the dependencies of the SHC response profile relative to shifting carrier concentrations, the SHC was evaluated as a continuous function of the chemical potential. By scanning the Fermi energy across a window of $\pm 1\text{ eV}$ around the E$_F$ with an explicit increment step of $0.01\text{ eV}$,  
the SHC tensor element was computed by integrating the spin Berry curvature over a dense, interpolated $\mathbf{k}$-point mesh of $150 \times 150 \times 150$. This ensures high resolution around avoided crossings in the electronic structure.

\begin{table}[t]
\caption{Lattice constants of pristine as well as doped systems. In the last two configurations, the void 2$a$ Wyckoff sites were substituted by Pt/Ta atoms. Iso. and non-iso. is abbreviation for isoelectronic and non-isoelectronic substitution, respectively. }
\label{tab:lattice_constants}
\begin{ruledtabular}
\begin{tabular}{lcc}
System & Configuration & $a$ (\AA) \\
\hline
CoAs$_3$~\cite{Mandel1971}         & ---       & 8.195 \\
CoSb$_3$~\cite{Rodrigues2022}      & ---       & 9.022 \\
RhAs$_3$~\cite{Kjekshus1974}       & ---       & 8.451 \\
RhSb$_3$~\cite{Kjekshus1974}       & ---       & 9.232 \\
Rh$_{1/2}$Ir$_{1/2}$Sb$_3$     & Iso.      & 9.384 \\
Rh$_{1/2}$Ir$_{1/2}$Sb$_{3/2}$Bi$_{3/2}$       & Iso.      & 9.563 \\
Rh$_{1/2}$Pt$_{1/2}$Sb$_3$     & Non-iso.  & 9.469 \\
Pt$_{1/4}$RhAs$_3$     & 2a        & 8.579 \\
Ta$_{1/4}$RhAs$_3$     & 2a        & 8.628 \\
\end{tabular}
\end{ruledtabular}
\end{table}

\section{RESULTS AND DISCUSSION}

\subsection{Electronic structure of pristine compounds}
\begin{figure*}[t]
\centering
\includegraphics[width=18 cm]{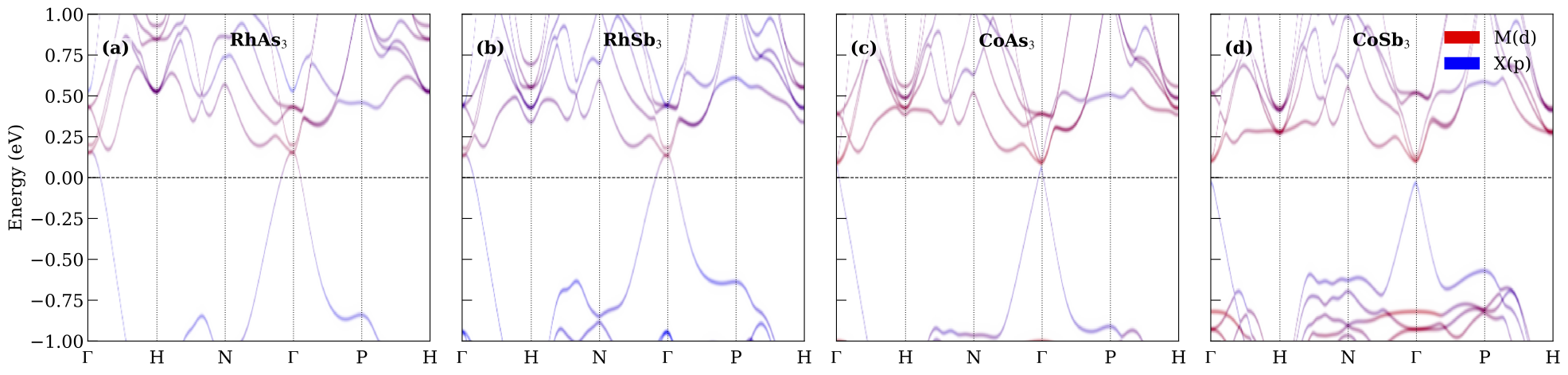}
\caption{Atom-projected \textit{ab initio} relativistic band structures of the MX\(_3\) family, where M = Rh or Co, and X = As or Sb. Except for CoSb\(_3\), all other members of the MX\(_3\) family exhibit semimetallic behavior. The VBM and CBM have predominant contributions from X-\(p\) and M-\(d\) orbitals.}
\label{fig:Bands}
\end{figure*}

Before exploring the spin transport, we systematically analyze the relativistic \textit{ab-initio} electronic structure of the four pristine skutterudite compounds with spin-orbit coupling included self-consistently, as shown in Fig.~\ref{fig:Bands}. Our calculations reveal a clear chemical trend governed by changing the transition metal and the pnictogen cage elements. While \(\text{CoSb}_3\) stabilizes as a true narrow-gap semiconductor, all other members of the MX\(_3\) family exhibit a semimetallic character characterized by a slight energetic overlap between the valence band maximum (VBM) and the conduction band minimum (CBM) at the \(\Gamma\) point. This observation aligns well with earlier electronic-structure calculations that detail the sensitivity of near-Fermi pseudogaps to pnictogen-cage modifications~\cite{Koga2005,Khan2021}.

A key topological feature is observed directly at the center of the Brillouin zone (i.e., the \(\Gamma\) point). Instead of a conventional linear Dirac dispersion typically cited in pseudo-scalar relativistic models~\cite{Smith2011}, the Rh-based compounds retain a predominantly parabolic low-energy dispersion; the Co-based systems exhibit a noticeably steeper valence-band branch that appears nearly linear over the energy range shown in the band structure plots. This suggests that the Co compounds may be farther from the idealized quadratic character of the VBM than the Rh compounds. This specific evolution from linear bands to a quadruply degenerate quadratic contact point upon the self-consistent inclusion of SOC is consistent with established theoretical insights into the low-energy physics of binary skutterudites, such as those presented by Smith \textit{et al.}~\cite{Smith2011} and Zhu \textit{et al.}~\cite{Zhu2018}.

Within the valence-band manifold shown in Fig.~\ref{fig:Bands}, extending from the valence-band maximum at \(0\text{ eV}\) down to \(-1.0\text{ eV}\), the state contributions arise primarily from pnictogen \(p\) orbitals with a smaller contribution from transition-metal \(d\) states. Conversely, in the low-energy conduction-band region extending from the E$_F$ to \(1.0\text{ eV}\), the states are dominated by transition-metal \(d\) orbitals, with a substantial but comparatively smaller contribution from As \(p\) orbitals. This profound mixing across the E$_F$ indicates that the near-gap states cannot be simplistically decoupled into unhybridized individual atomic manifolds, a characteristic that strongly mirrors the hybridized four-band models reported earlier~\cite{Koga2005,Khan2021}. The coexistence of these hybridized bands, the resulting multifold quadratic degeneracies at the zone center, and the sensitive semimetallic-to-semiconducting energy gaps provide an ideal physical environment to support a large spin Berry curvature.

\begin{figure*}[t]
\centering
\includegraphics[width=18 cm]{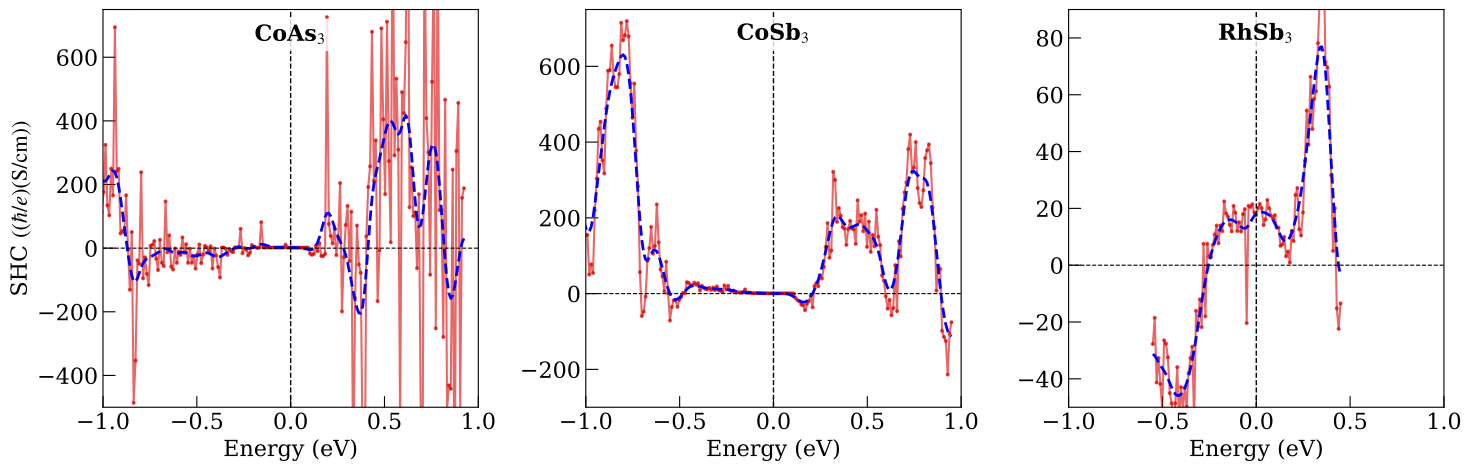}
\caption{
Intrinsic spin Hall conductivity as a function of the chemical potential for CoAs\(_3\), CoSb\(_3\), and RhSb\(_3\). The vertical dashed lines at zero eV indicate the E$_F$, and the horizontal dashed lines mark zero SHC. The red solid line and blue dashed line represent, the raw and smoothened interpolated data, respectively.
}
\label{fig:SHC_parent_compounds}
\end{figure*}

\subsection{Spin Hall Response of pristine materials}

\begin{figure}[t]
\centering
\includegraphics[width= 8.5cm, height=5.5 cm]{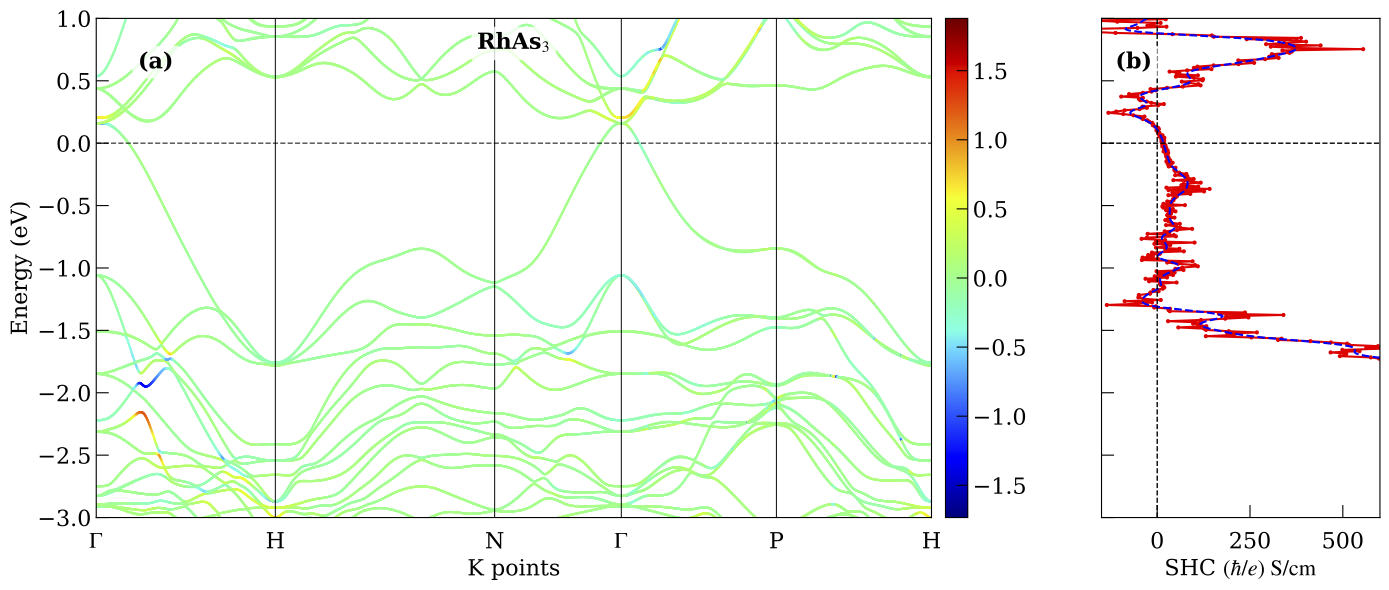}
\caption{
(a) Band structure of RhAs\(_3\) along the selected high-symmetry path, color-coded according to the momentum-resolved spin Berry curvature \(\log_{10}|\Omega_{xy}^{\mathrm{spin},z}(\mathbf{k})|\).
(b) Intrinsic spin Hall conductivity as a function of the chemical potential. The horizontal dashed line in panel (a) and the corresponding reference-energy line in panel (b) indicate the E$_F$.
}
\label{fig:RhAs3_kresolved_SHC}
\end{figure}

Motivated by these near-Fermi electronic features and the potential for pronounced SBC, we analyze the intrinsic SHC of the pristine compounds by obtaining WFTB models as explained in the methodology section. Our models are in excellent agreement (not shown) with the relativistic \textit{ab initio} band structures. 
Calculated SHCs of CoAs$_3$, CoSb$_3$, and RhSb$_3$ are shown in Fig.~\ref{fig:SHC_parent_compounds}, while that of RhAs$_3$ is presented together with its momentum-resolved SBC in Fig.~\ref{fig:RhAs3_kresolved_SHC}.
At the E$_F$, all four compounds exhibit a significantly suppressed SHC response. 

Initially, one might assume that this weak response means the material lacks spin Berry curvature entirely. However, the momentum-resolved spin Berry curvature plot for RhAs$_3$ in Fig.~\ref{fig:RhAs3_kresolved_SHC} reveals highly localized hot spots residing along the high-symmetry paths, particularly concentrated near the Brillouin zone center (\(\Gamma\) point). One example of such opposing contributions to spin Berry curvature can be seen around \(-2\text{ eV}\) along the \(\Gamma\)--H direction in the BZ. This apparent suppression of the macroscopic SHC may stem from substantial cancellation of the positive and negative spin Berry curvature contributions from these localized hot spots, which perfectly cancel each other out when integrated across the occupied momentum space.

\begin{figure*}[t]
\centering
\includegraphics[width=17.5 cm]{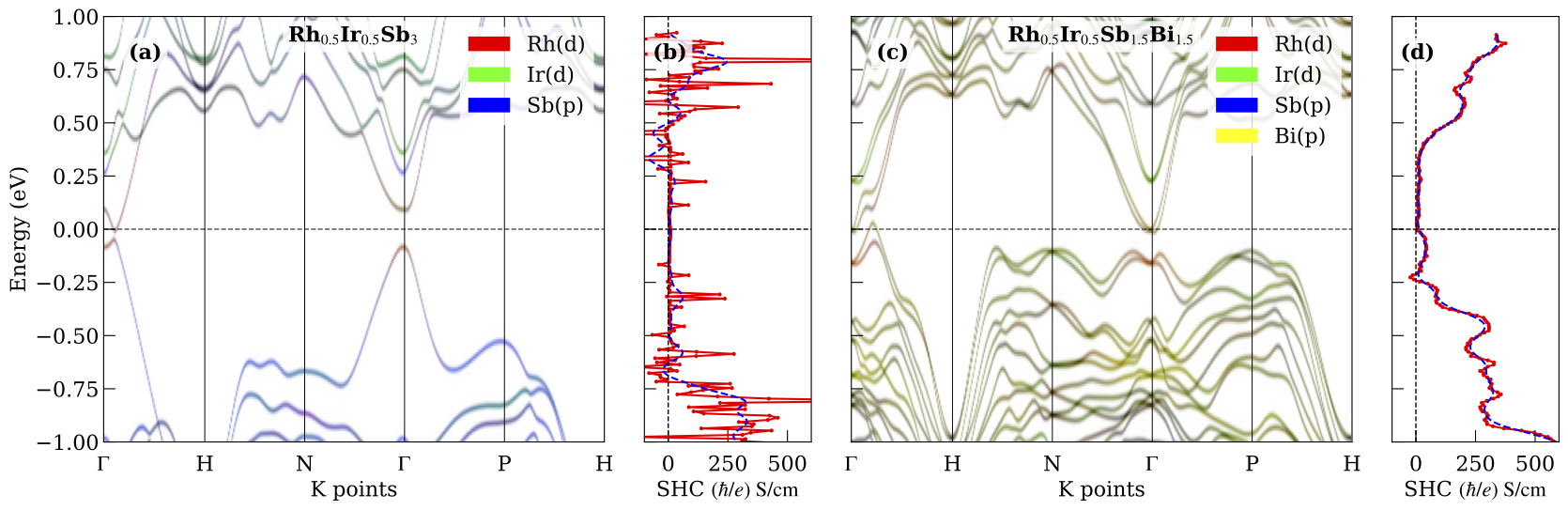}
\caption{
Orbital-resolved band structures and intrinsic SHC of the iso-electronically substituted systems. Panels (a) and (b) show the orbital-projected band structure and SHC, respectively, for Rh\(_{0.5}\)Ir\(_{0.5}\)Sb\(_3\), while panels (c) and (d) show the corresponding results for Rh\(_{0.5}\)Ir\(_{0.5}\)Sb\(_{1.5}\)Bi\(_{1.5}\). The horizontal dashed lines denote the E$_F$. The orbital contributions from different atoms are marked by different colors.
}
\label{fig:isoelectronic_fatband_SHC}
\end{figure*}

As the chemical potential is systematically scanned away from the E$_F$ into the adjacent valence or conduction blocks, the system rapidly moves past this cancellation zone, tapping directly into these dense pockets of spin Berry curvature, resulting in a large SHC. This mechanism aligned well with the established physics of intrinsic spin Hall behavior in narrow-gap and topological semimetallic architectures, where a substantial local SHC remains hidden until activated via deliberate carrier engineering or chemical substitution~\cite{Guo2008,Sun2016}.

\subsection{Spin Hall response under isoelectronic heavy element substitution}


Having established that pristine skutterudites suffer from a global cancellation of spin Berry curvature at \(E_F\), we next examine whether chemical substitution with heavier elements can perturb the band topology and unmask some better responses. To systematically manipulate the relativistic electronic profile and maximize the SBC, we first explore a targeted chemical substitution strategy. Among the four materials investigated, \(\text{RhSb}_3\) was chosen as the ideal host architecture, as \(\text{RhSb}_3\) already possesses the heaviest constituent elements in the pristine series, and hence a larger SOC strength. Moreover, \(\text{RhSb}_3\) stabilizes on the brink of a topological phase boundary~\cite{Koga2005},
exhibiting a marginal semimetallic band overlap or zero-gap semiconducting nature. This critical state makes its low-energy bands exceptionally sensitive to relativistic perturbations. 

\begin{figure}[t]
\centering
\includegraphics[width=8.5cm]
{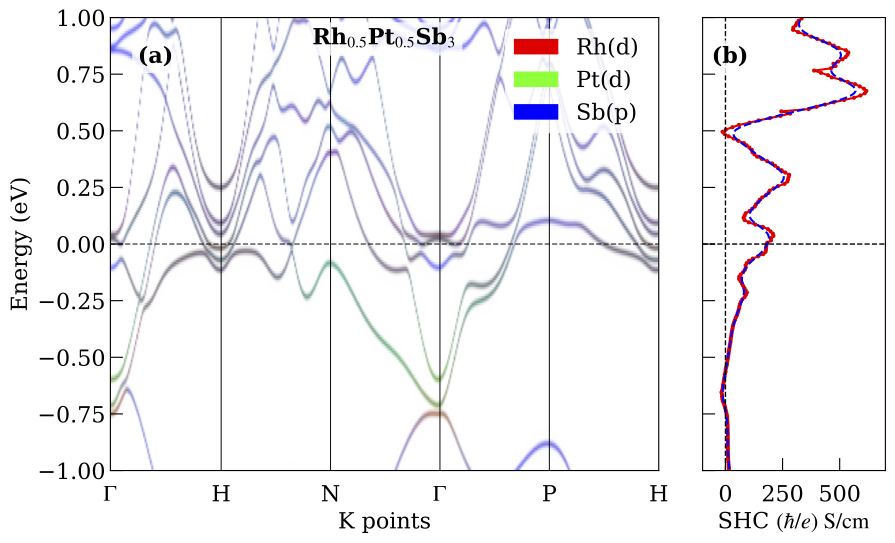}
\caption{
(a) Orbital-resolved band structure and (b) intrinsic SHC of Rh$_{0.5}$Pt$_{0.5}$Sb$_3$. Different orbital character are color coded. The horizontal dashed line marks the E$_F$.
}
\label{fig:nonisoelectronic_fatband_SHC}
\end{figure}

Based on these considerations, we modeled a 50\% isoelectronic substitution of the transition metal site, replacing \(\text{Rh}\) with its heavier counterpart \(\text{Ir}\) to form Rh\(_{0.5}\)Ir\(_{0.5}\)Sb\(_3\). The calculated relativistic electronic band structure is shown in Fig.~\ref{fig:isoelectronic_fatband_SHC}(a). The introduction of the heavy $5d$ element dramatically amplifies the spin-orbit coupling strength. As observed along the high-symmetry paths in Fig.~\ref{fig:isoelectronic_fatband_SHC}(a), this enhanced relativistic interaction lifts the remaining multifold degeneracies at the zone center. Instead of generating a network of symmetry-protected crossings at \(E_F\), the amplified SOC pushes the valence and conduction bands apart, opening a clean semiconducting energy gap. This SOC-induced gap expansion directly modifies the energy-dependent SHC scan shown in Fig.~\ref{fig:isoelectronic_fatband_SHC}(b). While stronger SOC brings some finite peaks close to E$_F$, the SHC drops to zero exactly at the E$_F$. This suppression signals the formation of a structurally driven electronic gap.  Hence,  
these results indicate that isoelectronic substitution at the metal site alone is insufficient to enhance the material's intrinsic spin response.

To test if a heavier pnictogen could fix the gap or create a useful band inversion, we replaced 50\% of Sb with Bi to get \(\text{Rh}_{0.5}\text{Ir}_{0.5}\text{Sb}_{1.5}\text{Bi}_{1.5}\) [Fig.~\ref{fig:isoelectronic_fatband_SHC}(c,d)]. The modified band structure shows that this heavy element substitution restores a semimetallic state, with some valence bands crossing above the E$_F$ and forming hole pockets. However, despite these itinerant carriers, the SHC near the E$_F$ remains suppressed. This happens because Bi's strong atomic spin-orbit coupling pushes the regions responsible for finite SBC, far away from the Fermi energy into deeper bands. The bands that do cross the E$_F$ are highly dispersive and lack the topological features needed to generate strong curvature peaks. Moreover, any small curvature contributions that do arise in these pockets cancel out when integrated over the entire Brillouin zone due to sign differences. So, while heavy-element substitution can restore metallicity, it fails to align the spin-active hot spots with the E$_F$, showing that isoelectronic substitutions alone are not enough to optimize spin-current generation in these materials.

This fundamental limitation highlights a key insight: because isoelectronic substitution preserves the total valence electron count, it cannot shift the chemical potential relative to the SOC-modified band features. To break this impasse and position SBC hot spots precisely at the E$_F$, non-isoelectronic carrier tuning and site-selective doping strategies are explored in the following section.

\subsection{Role of Site-selective doping for enhanced spin response}

Based on our results in the previous sections, a successful design strategy must accomplish two goals simultaneously: (i) alter the valence electron count (non-isoelectronic carrier tuning) to shift \(E_F\), and (ii) introduce strong spin-orbit coupling (SOC) while breaking local structural symmetry to unmask suppressed SBC hot spots. Here, we systematically evaluate non-isoelectronic doping, contrasting framework substitution at the transition-metal \(M\)-site with interstitial filling at the vacant $2a$ Wyckoff cage sites.

We first explore \(M\)-site non-isoelectronic substitution by replacing 50\% of Rh (at 8$c$ Wycoff positions) with Pt forming \(\mathrm{Rh}_{0.5}\mathrm{Pt}_{0.5}\mathrm{Sb}_3\). This non-isoelectronic substitution of Pt (\(5d^9\)) introduces an extra electron carriers compared to Rh (\(4d^8\)) while simultaneously increasing the relativistic SOC strength as we move from 4$d$ to 5$d$ element. As shown in the orbital-resolved band structure and SHC plot in Fig.~\ref{fig:nonisoelectronic_fatband_SHC}(a) and (b), the addition of Pt $5d$ electrons shifts the chemical potential upward into a band-crossing manifold rich in SOC-induced splittings. Consequently, \(\mathrm{Rh}_{0.5}\mathrm{Pt}_{0.5}\mathrm{Sb}_3\) exhibits a finite, non-zero SHC of approximately \(200~(\hbar/e)(\Omega\cdot\mathrm{cm})^{-1}\) directly at the E$_F$. This result confirms that non-isoelectronic doping successfully overcomes the complete SHC suppression found in pristine and iso-electronically substituted compounds. However, whether such substitutional doping at 8$c$ Rh sites is the only route to enhance SHC remained a natural question.

\begin{figure*}[t]
\centering
\includegraphics[width=\textwidth]{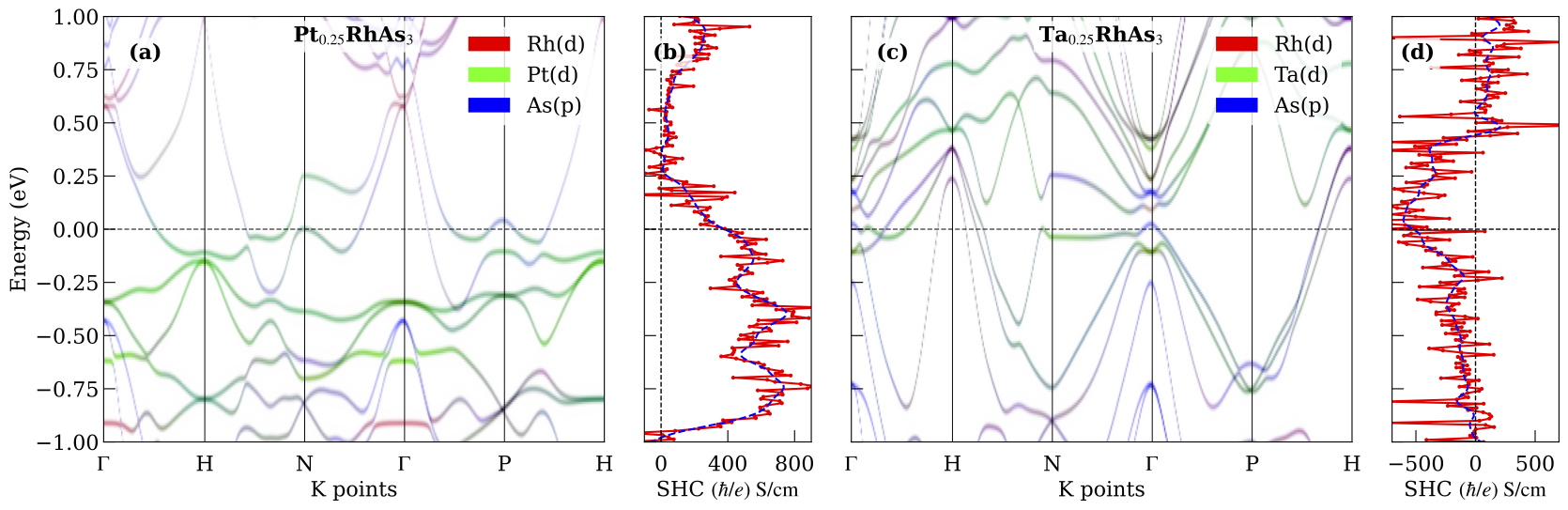}
\caption{
Orbital-resolved band structures and intrinsic spin Hall conductivities for substitution at the $2a$ site. Panels (a) and (b) show the orbital-projected band structure and SHC, respectively, of Pt\(_{0.25}\)RhAs\(_3\), whereas panels (c) and (d) show the corresponding results for Ta\(_{0.25}\)RhAs\(_3\). The orbital projections include Rh \(d\), dopant (Pt or Ta) \(d\), and As \(p\) contributions. The horizontal dashed lines indicate the E$_F$.
}
\label{fig:2a_fatband_SHC}
\end{figure*}

To answer the above question, we turn to filling the intrinsic crystallographic voids at the $2a$ Wyckoff positions of the skutterudite structure with heavy $5d$ elements, viz-a-viz, Pt and Ta. Full variable-cell relaxation for interstitial Pt filling in \(\mathrm{RhAs}_3\) (\(\mathrm{Pt}_{0.25}\mathrm{RhAs}_3\)) reveals a stable structural expansion: the nearest-neighbor \(\mathrm{Rh}-\mathrm{As}\) bond distance expands slightly to \(2.462\text{ \AA}\) (compared to the pristine value of \(2.434\text{ \AA}\) in \(\mathrm{RhAs}_3\)),  accommodating the bulky guest Pt atom with optimized interatomic distances of \(3.715\text{ \AA}\) for \(\mathrm{Pt}-\mathrm{Rh}\) and \(3.193\text{ \AA}\) for \(\mathrm{Pt}-\mathrm{As}\). As shown in Fig.~\ref{fig:2a_fatband_SHC}(a) and (b), \(\mathrm{Pt}_{0.25}\mathrm{RhAs}_3\) exhibits a dramatic boost in intrinsic spin transport, reaching an SHC of approximately \(400~(\hbar/e)(\Omega\cdot\mathrm{cm})^{-1}\) directly at \(E_F\)-doubling the response achieved via \(M\)-site substitution. Furthermore, scanning the chemical potential slightly below \(E_F\) traverses through dense, constructive SBC hot spots, driving the SHC to peak values of \(650-700~(\hbar/e)(\Omega\cdot\mathrm{cm})^{-1}\), whereas scanning above \(E_F\) leads to intra-band cancellations that suppress the signal.

\begin{figure*}[t]
\centering
\includegraphics[width=17.5 cm, height=9.5 cm]{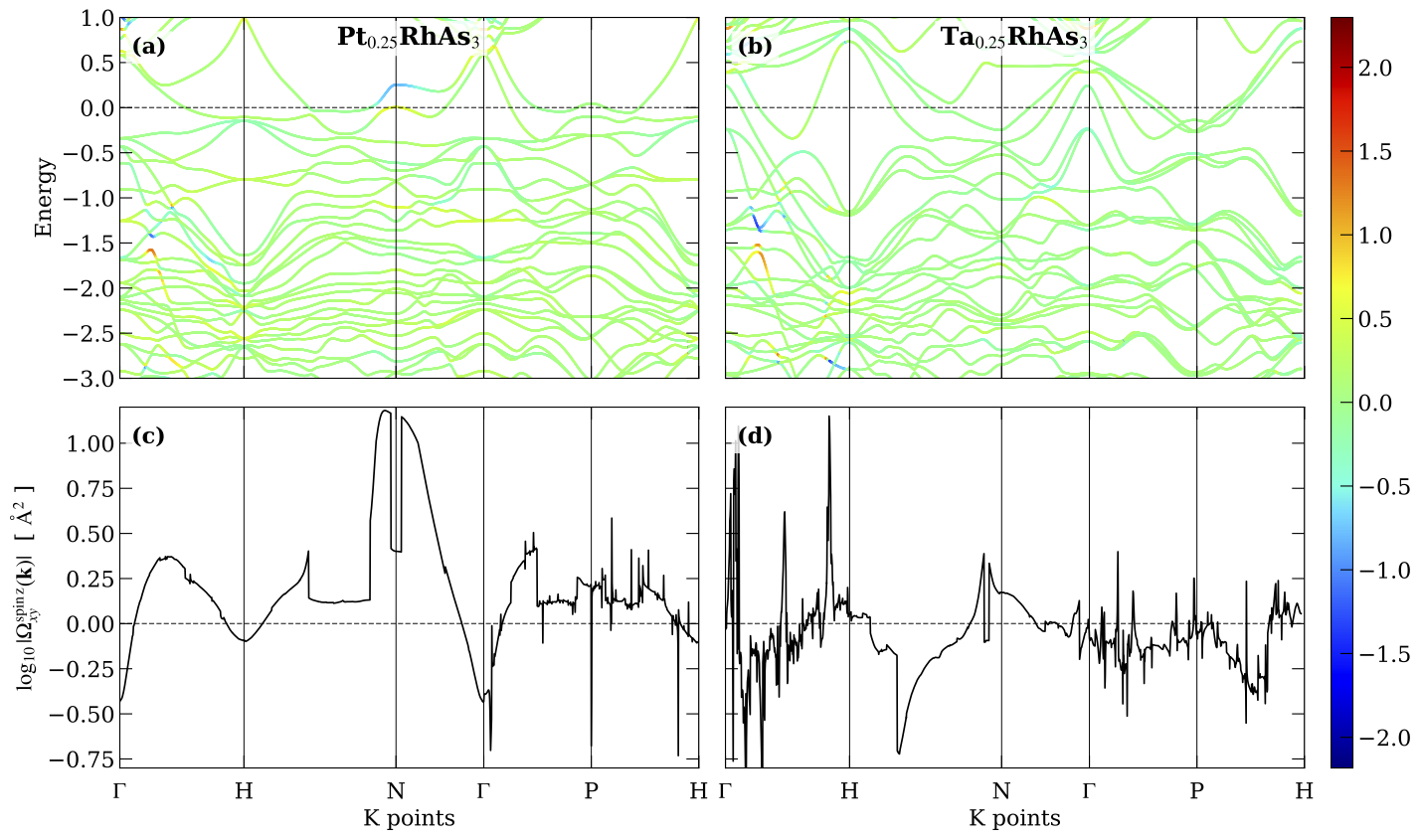}
\caption{
Panels (a) and (b) show the band structures interstitial substitution of Pt and Ta at 2$a$ i.e. Pt\(_{0.25}\)RhAs\(_3\) and Ta\(_{0.25}\)RhAs\(_3\), respectively, color-coded by \(\log_{10}|\Omega_{xy}^{\mathrm{spin},z}(\mathbf{k})|\). Panels (c) and (d) show the corresponding SBC along the selected high-symmetry path. The horizontal dashed lines denote the E$_F$.
}
\label{fig:2a_spin_Berry_curvature}
\end{figure*}

The microscopic origin of this enhanced response is illustrated by mapping the momentum-resolved spin Berry curvature in Fig.~\ref{fig:2a_spin_Berry_curvature}(a,c). While large SHC of elemental fcc Pt arises from SOC-induced avoided crossings near the \(L\) and \(X\) high-symmetry points~\cite{Guo2008}, the bcc-like structural environment of the filled skutterudite host redistributes these SBC features across momentum space. In \(\mathrm{Pt}_{0.25}\mathrm{RhAs}_3\), the SBC is intensely concentrated near SOC-induced avoided crossings, with a dominant hot spot localized near the \(N\) point along the high-symmetry path as seen in Fig.~\ref{fig:2a_spin_Berry_curvature}(c). Analysis of the relativistic orbital projections in Fig.~\ref{fig:2a_fatband_SHC}(a) confirms that the itinerant states near \(E_F\) are derived from interstitial Pt $5d$ orbitals. 

To clarify whether this SHC enhancement is universally driven by any heavy $5d$ interstitial element or requires a specific valence electron count, we contrast \(\mathrm{Pt}_{0.25}\mathrm{RhAs}_3\) with $2a$-site Ta filling (\(\mathrm{Ta}_{0.25}\mathrm{RhAs}_3\)), presented in Figs.~\ref{fig:2a_fatband_SHC}(c,d) and \ref{fig:2a_spin_Berry_curvature}(b,d). Interestingly, Ta filling also generates a remarkably large SHC magnitude exceeding \(400~(\hbar/e)(\Omega\cdot\mathrm{cm})^{-1}\) at the E$_F$ as shown in Fig.~\ref{fig:2a_fatband_SHC}(d). However, because Ta possesses an electron-deficient \(5d^3\) configuration compared to Pt's \(5d^9\), \(E_F\) shifts to intersect a different region of the relativistic band structure where negative SBC hot spots dominate as shown in Fig.~\ref{fig:2a_spin_Berry_curvature}(b,d). Consequently, the resulting SHC in Ta based system undergoes a complete sign reversal relative to the Pt-doped system.

Having established the strong site dependence of the intrinsic spin Hall response, we next examine the corresponding SNC, which provides the thermal counterpart of the electrically driven spin Hall effect. Since the SNC depends on the energy variation of the SHC within the thermal window around the chemical potential [Eq.~(\ref{eq:snc})], a large SHC does not necessarily imply a proportionally large SNC.

\begin{figure*}[t]
\centering
\includegraphics[width=12 cm, height= 5 cm]{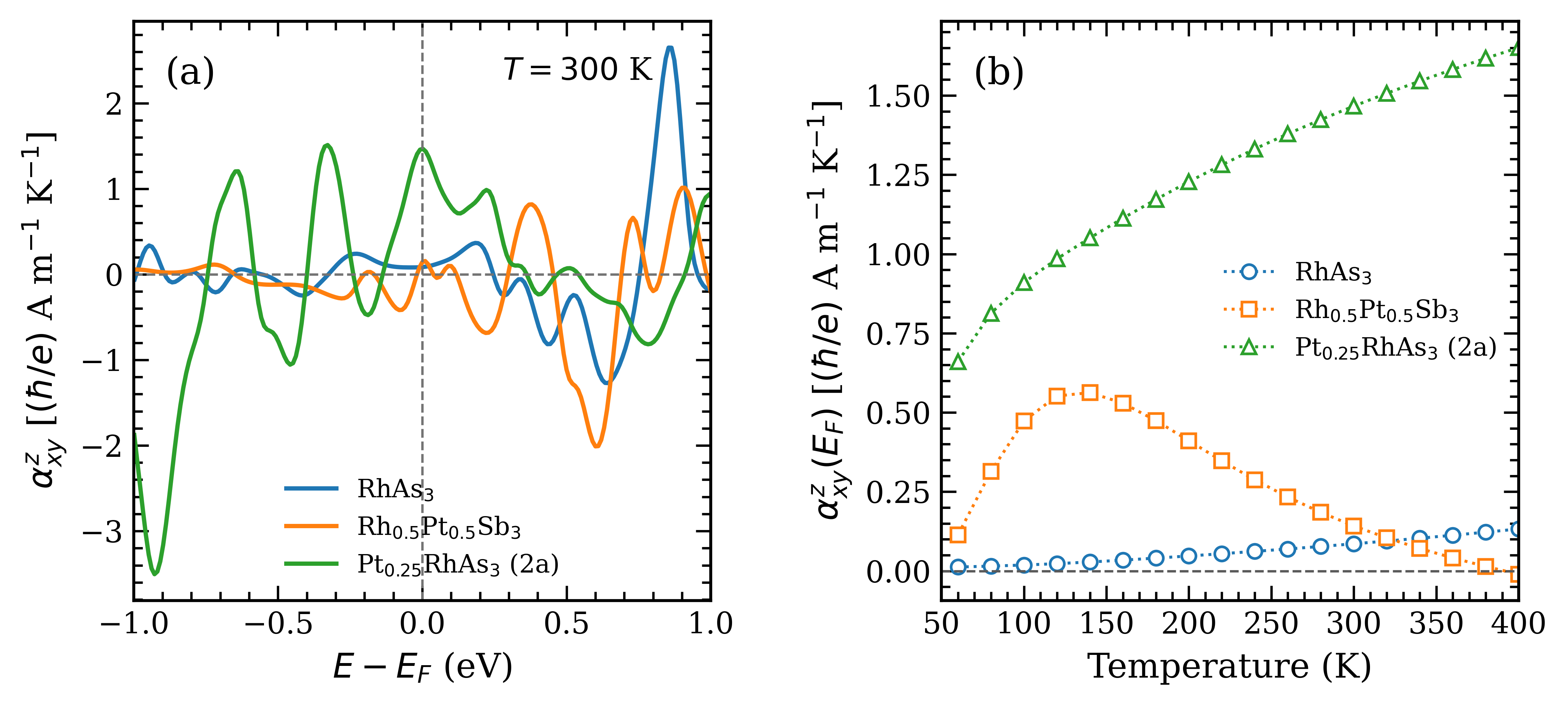}
\caption{(a) Intrinsic spin Nernst conductivity $\alpha_{xy}^{z}$ as a function of energy relative to the E$_F$ at $T=300$ K for pristine RhAs$_3$, $M$-site substituted Rh$_{0.5}$Pt$_{0.5}$Sb$_3$, and interstitially filled Pt$_{0.25}$RhAs$_3$. (b) Temperature dependence of $\alpha_{xy}^{z}$ evaluated at the E$_F$ for the same systems.}
\label{fig:SNC}
\end{figure*}

Figure~\ref{fig:SNC}(a) compares the SNC of pristine RhAs$_3$, $M$-site Pt-substituted Rh$_{0.5}$Pt$_{0.5}$Sb$_3$, and interstitially filled Pt$_{0.25}$RhAs$_3$ at $T=300$ K. The SNC exhibits a pronounced dependence on the chemical potential, including several sign changes, reflecting the strong energy dependence of the underlying spin Hall conductivity. At the E$_F$, pristine RhAs$_3$ exhibits a relatively small SNC of approximately $0.09~(\hbar/e)\,\mathrm{A\,m^{-1}K^{-1}}$, which increases to approximately $0.14~(\hbar/e)\,\mathrm{A\,m^{-1}K^{-1}}$ for Rh$_{0.5}$Pt$_{0.5}$Sb$_3$. In contrast, interstitial Pt filling gives a much larger value of approximately $1.46~(\hbar/e)\,\mathrm{A\,m^{-1}K^{-1}}$ in Pt$_{0.25}$RhAs$_3$.
The temperature dependence shown in Fig.~\ref{fig:SNC}(b) further highlights the distinct effect of interstitial Pt filling. For Pt$_{0.25}$RhAs$_3$, the SNC increases from approximately $0.66~(\hbar/e)\,\mathrm{A\,m^{-1}K^{-1}}$ at $60$ K to $1.47~(\hbar/e)\,\mathrm{A\,m^{-1}K^{-1}}$ at $300$ K and remains large up to $400$ K. By contrast, Rh$_{0.5}$Pt$_{0.5}$Sb$_3$ exhibits a maximum SNC of approximately $0.56~(\hbar/e)\,\mathrm{A\,m^{-1}K^{-1}}$ around $140$ K, followed by a substantial decrease with increasing temperature. The pristine RhAs$_3$ response remains comparatively small throughout the investigated temperature range.

These results demonstrate that the location of the Pt dopant plays a particularly important role in the thermally driven spin response. 
Although $M$-site Pt substitution already enhances the spin Hall response, placing Pt at the vacant $2a$ cage site enhances the room-temperature SNC by approximately one order of magnitude compared with the $M$-site-substituted system.
This indicates that strong SOC alone is not sufficient to account for the enhancement; rather, the site-dependent modification of the relativistic electronic structure and the associated energy variation of the SBC near $E_F$ are crucial for generating a large SNC.

Experimentally, the synthesis of these interstitially filled and substituted skutterudite crystal structures might be feasible using advanced growth techniques such as high-pressure synthesis or flux-assisted crystal growth commonly employed for such compounds. Validating these predicted phases experimentally will establish site-selective spin response  as a viable design strategy for room-temperature spin-current generation.

\section{SUMMARY}

In summary, we have investigated the intrinsic spin transport in binary skutterudites and established a site-selective cage engineering strategy to optimize their performance. While pristine compounds and isoelectronic alloys exhibit a strongly suppressed net spin Hall response due to momentum-space cancellation of spin Berry curvature, non-isoelectronic \(M\)-site substitution (\(\mathrm{Rh}_{0.5}\mathrm{Pt}_{0.5}\mathrm{Sb}_3\)) activates a moderate response of \(\sim 200~(\hbar/e)(\Omega\cdot\mathrm{cm})^{-1}\). In contrast, interstitial filling at vacant $2a$ cage sites with heavy $5d$ elements (\(\mathrm{Pt}\), \(\mathrm{Ta}\)) unmask dense SBC hot spots near the E$_F$. This elevates the intrinsic spin Hall conductivity to over \(400~(\hbar/e)(\Omega\cdot\mathrm{cm})^{-1}\) at \(E_F\) (and up to \(650-700~(\hbar/e)(\Omega\cdot\mathrm{cm})^{-1}\) upon chemical-potential tuning), while early (\(\mathrm{Ta}\)) versus late (\(\mathrm{Pt}\)) transition-metal filling grants robust sign-reversibility over the generated spin current. 
The site dependence is even more pronounced for the thermally driven SNC. At $300$ K, Pt$_{0.25}$RhAs$_3$ exhibits an SNC of approximately $1.46~(\hbar/e)\,\mathrm{A\,m^{-1}K^{-1}}$ at the E$_F$, compared with approximately $0.14~(\hbar/e)\,\mathrm{A\,m^{-1}K^{-1}}$ for $M$-site Pt substitution. Moreover, the SNC of the interstitially filled system remains large up to $400$ K, demonstrating a robust room-temperature SNC response.
These findings highlight interstitial cage filling as a versatile design principle for high-performance spintronic and spin-caloritronic materials.

\begin{acknowledgments}
SD acknowledges National Supercomputing Mission (NSM) for providing computing resources of ‘PARAM Kamrupa’ at IIT Guwahati, which is implemented by C-DAC and supported by the Ministry of Electronics and Information Technology (MeitY) and Department of Science and Technology (DST), Government of India. S.K.P. acknowledges the NFSG grant from BITS-Pilani, Dubai campus, which supported this research.
\end{acknowledgments}

\end{document}